# Harnessing Infant Cry for swift, cost-effective Diagnosis of Perinatal Asphyxia in low-resource settings

Charles C. Onu

*Abstract*— Perinatal Asphyxia is one of the top three causes of infant mortality in developing countries, resulting to the death of about 1.2 million newborns every year. At its early stages, the presence of asphyxia cannot be conclusively determined visually or via physical examination, but by medical diagnosis. In resource-poor settings, where skilled attendance at birth is a luxury, most cases only get detected when the damaging consequences begin to manifest or worse still, after death of the affected infant. In this project, we explored the approach of machine learning in developing a low-cost diagnostic solution. We designed a support vector machine-based pattern recognition system that models patterns in the cries of known asphyxiating infants (and normal infants) and then uses the developed model for classification of `new' infants as having asphyxia or not. Our prototype has been tested in a laboratory setting to give prediction accuracy of up to 88.85%. If higher accuracies can be obtained, this research may be a key contributor to the 4th Millennium Development Goal (MDG) of reducing mortality in under-five children.

## I. INTRODUCTION

Perinatal Asphyxia is a condition that results when a newborn fails to establish proper respiration immediately after birth, leading to shortage in supply of oxygen to the brain. Every year, about 1.2 million infants die from perinatal asphyxia [1] and about an equal number suffer severe life-long conditions such as cerebral palsy, deafness, and different degrees of damage to the Central Nervous System (CNS).

Globally, the mortality rate in under-five children dropped from 12.4 million in 1990 to 6.6 million in 2012 [2], indicating significant progress at achieving the 4[th] Millennium Development Goal (MDG). However, of recent concern is the rising proportion of infant deaths occurring within the neonatal (first month after birth) period, which currently accounts for a whooping 4 million annually. In developing countries, perinatal asphyxia is a leading cause of morbidity and mortality of infants within this phase [3].

At its early stages, the presence of asphyxia cannot be conclusively determined visually or via physical examination but by medical diagnosis (involving blood sampling and series of tests). In resource-poor settings, where skilled attendance at birth is a luxury, most cases only get detected when the damaging consequences begin to emerge or worse still, after death of the affected infant. Our objective in this project was thus to develop a diagnostic solution that would not only enable the timely recognition of perinatal asphyxia in newborns but also be a cost-efficient alternative for the developing world. We explored the approach of machine learning, designing a support vector machine-based pattern recognition system that models patterns in the cries of known asphyxiating infants (and normal infants) and then uses the developed model for classification of "new" infants as having asphyxia or not.

Amongst efforts geared at solving this problem, the works of Reyes-Galaviz O. F. and Reyes-Garcia C. A. [4] has been prominent. They emphasized the crucial importance of early diagnosis of pathologies like asphyxia in newly born babies and went ahead to develop a system that processes infant cry to automatically recognize babies born with asphyxia using Neural Networks. Their work was based on the fact that "crying in babies is a primary communication function, governed directly by the brain, and any alteration on the normal functioning of the babies' body is reflected in the cry."[4] In developing the system, they collected cry samples of normal, deaf and asphyxiating babies into a corpus (the *Baby Chillanto Database*) and applied the techniques of automatic speech recognition to create a pattern recognition model. Their experiments yielded classification precision of up to 86%.

Leveraging on the experience of Reyes-Galaviz and Reyes-Garcia, we experimented using Support Vector Machines (SVMs), for performance comparison and to pursue higher classification accuracies; given the knowledge that SVMs provide a very good out-of-sample performance and scale well on speech recognition problems [5]. To develop this system, the *Baby Chillanto Database* was obtained courtesy of the *National Institute of Astrophysics and Optical Electronics, CONACYT, Mexico*. Of interest to our research were the 1049 normal and 340 asphyxia cry samples contained therein (and separated by us in the ratio 60:20:20 for training, cross-validation and testing, respectively).

Using MATLAB, each cry sample went through several signal processing stages; at the end of which feature vectors were extracted as coefficients of the Mel Frequency Cepstrum (MFC) and then used as input to the learning algorithm. Experiments were performed using two different types of Support Vector Machine Kernels – Polynomial Kernel and Radial Basis Function (RBF) Kernel. We report our procedures and results which show best classification accuracy of 88.85% obtained using the Polynomial Kernel.

In the next sections, we briefly describe the learning algorithm used, then discuss the approach in detail, present experimental results, and make conclusions and notes for improvement.

Charles C. Onu is an Electronics and Computer Engineer with the Fisher Foundation for Sustainable Development in Africa, Owerri, Nigeria (phone: +2348068864622; e-mail: onucharles@gmail.com).

## II. SUPPORT VECTOR MACHINES

A Support Vector Machine (SVM) is a state-of-the-art learning algorithm which operates mainly on the principle of distance, by monitoring similarities between features of samples in a dataset. A key decision in designing an SVM learning system lies in choosing a Kernel function that is right for the problem set. A kernel function is a similarity function that defines the basis for measurement of the proximity of two or a combination of samples from a dataset. Not all similarity functions make valid kernels. A valid kernel must satisfy Mercer's theorem [6].

There are several kernel functions. The two (2) used in this experiment are described below:

### A. Radial Basis Function (RBF) Kernel

RBF is one of the most popular kernels in use and is very suited for majority of applications. It nonlinearly maps samples into a higher dimensional space so it, can handle the case when the relation between class labels and attributes is non-linear [7].

$$K(x,y) = \exp(\gamma \|x - y\|_2^2) \tag{1}$$

### B. Polynomial Kernel

The polynomial Kernel also allows the learning of non-linear models, as it looks not only at the given features of input samples to determine their similarity but also combinations of these features [8].

$$K(x,y) = (X^T + c)^d \tag{2}$$

## III. METHODOLOGY

The Baby Chillanto dataset, which was used for the experiments, consists of 1049 normal and 340 asphyxia audio samples, subsequently tagged as negative and positive samples respectively. Each audio sample was provided as a 1-second (.wav) recording of infant cry. In order to perform recognition, we designed a machine learning pipeline of five (5) distinct phases: Audio Sampling, Feature Extraction, Feature Scaling, Training/Cross-Validation and Testing. The first 3 stages focus on processing the signal and preparing the samples, while the last two stages cover the actual pattern recognition system. MATLAB was used in writing the code for all the stages of the pipeline.

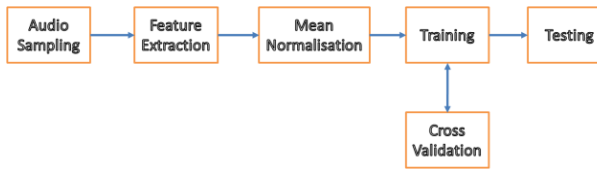

Fig. 1: Block diagram of Recognition Process

**Signal Processing Phase**

### A. Audio Sampling

All the negative and positive audio files were read into MATLAB. Though each sample has a duration of one second, the resulting vectors of the samples were of varying length due to different bit rates. Thus, in order to ensure that the resultant vectors were of equal dimension as this is a requirement for learning from data, 128 kilo-bits of data (corresponding to the lowest bit rate among all samples) was extracted from every audio sample. At a constant sample rate of 16bits/sample, this audio sampling phase resulted to an 8000 by 1vector for each sample in the dataset.

### B. Feature Extraction

In this work, features were extracted as coefficients of the Mel Frequency Cepstrum (MFC). MFC coefficients are widely used in automatic speech recognition problems as they provide a representation of audio signals that closely mimic the human auditory system. It takes human perception sensitivity with respect to frequencies into consideration and thus is most appropriate for voice recognition [9]. Mathematically, the MFC coefficient is defined as:

$$c(n) = DCT(\log(|FFT(s(n))|)) \tag{3}$$

where, s(n) is the original signal at frame n, after application of pre-filtering and some windowing method.

Using Voicebox [10], a MATLAB toolbox for speech processing, MFC coefficients was computed at a rate of 44200Hz, resulting in 16 by 12 matrices for each sample. Thus, reducing the originally sampled data from 8000 features to 168 features, by rolling each matrix into a vector.

### C. Feature Scaling (Mean Normalization)

The standard deviation of the original set of MFC feature vectors was computed to be 2.1043 which is undesirable. Feature scaling is a process of normalizing a dataset so that every value is within a small, defined range, aiding quicker convergence of the learning algorithm.

Mean Normalization was thus taken to regulate the features in the dataset to a mean of 0 and standard deviation of approximately, 1 (actually 1.0027). After scaling, the values of each feature were approximately between −0.5 and +0.5. The equation for mean normalization is given as:

$$X_{norm} = \frac{X - \mu}{std} \tag{4}$$

The output of feature scaling using mean normalization includes the normalized data, a vector containing the original mean of each feature, and the standard deviation of the data set.

**Pattern Recognition Phase**

The mean-normalized MFC feature vectors for both negative and positive samples served as input to this phase. The dataset was divided into training (60%), cross-validation (20%) and test (20%) sets. For the 1049 negative samples, this gave: 630 training, 209 cross validation and 210 test samples; while the 340 positive samples resulted to: 204 training, 68 cross-validation and 68 test samples. Both sets (negative and positive) were then mixed and shuffled together, to give the final dataset for the pattern recognition process which contained 834 training, 277 cross-validation and 278 test set samples.

The training set was used in conjunction with the cross-validation set to obtain best values for the kernel parameters. This was achieved by running the learning algorithm severally on the training set using all possible combinations of the parameters to fit, testing the models generated on the cross-validation set and then, selecting the parameters of the model that gave the lowest prediction error. The fitted parameters were then used to train a combination of both the training and cross-validation sets to develop the final model that was used for classification in the testing phase.

|  | Training | Cross Validation | Test |
|---|---|---|---|
| Negative Samples | 630 | 209 | 210 |
| Positive Samples | 207 | 68 | 68 |
| Total | 834 | 277 | 278 |

Table 1: Distribution of Samples for Pattern Recognition phase

*D. Training/Cross-validation*

LIBSVM [11] was used to implement Support Vector Machine learning. Experiments were performed on two different SVM kernels – Polynomial and Radial Basis Function (RBF) Kernels.

Concretely, in the polynomial kernel experiment, the cross-validation set was used to get the best values for the degree of polynomial function, d and gamma, g; while, the constant k which represents the trade-off between higher and lower order terms was kept at 0 (homogeneous). The parameters were tested in the ranges:

**Degree of polynomial (d)**: [1, 2, 3, 4, 5, 6, 7, 8]

**Gamma (g):** [0.0060  0.0179  0.0536  0.1607  0.4821  1.4464  4.3393  13.0179]

Listing 1: Range of values used for cross-validation in polynomial kernel experiment

Note: Gamma was increased in multiples of 3, starting from the reciprocal of the number of features in the data set (1 / num_features).

After testing every possible combination of d and g – a total of 64 cross-validation tests were carried out – the best values of 1 and 0.0060, respectively, was obtained and then entered into the learning algorithm. The algorithm was then used to train the final training set, which was a combination of both the initial training and cross-validation. A similar experiment was carried out when using RBF kernel. In this case, the best values of the parameters to fit – regularization cost, C and gamma, g – were obtained as 1 and 0.00595, respectively.

**Gamma (g)**: [0.0060  0.0179  0.0536  0.1607  0.4821  1.4464  4.3393  13.0179]

**Regularization Cost (C)**: [0.01 0.03 0.1 0.3 1 3 10 30]

Listing 2: Range of values used for cross-validation in radial basis function kernel experiment

*E. Testing*

In order to ensure an objective classification and to measure how well the trained model generalizes on data other than that used to train, 20% of the samples were reserved for testing only. Each test sample – just as the training samples – went through the signal processing phase of audio sampling, feature extraction and feature scaling. Eventually, the test data, along with its label vector and the trained model, was passed to the classification algorithm.

**Error Metric**

Due to the fact that the dataset is significantly skewed – more negative than positive samples – it is important to use an evaluation metric which takes into consideration the unbalanced nature of the data, when reporting the accuracy of the system. Therefore, in addition to the average accuracy (the number of correctly classified samples divided by the total number of samples), the F-Score was used as an additional error metric in providing a deeper insight into the results. It is given as:

$$Precision, P = \frac{no\ of\ true\ positives}{no\ of\ predicted\ positives} \quad (5)$$

$$Recall, R = \frac{no\ of\ true\ positives}{no\ of\ actual\ positives} \quad (6)$$

$$F-Score = 2\frac{PR}{(P+R)} \quad (7)$$

## IV. RESULT

The results are presented in two parts:

*A. Polynomial kernel Experiment*

Training converged after 404 iterations with a training set error of 0.0162. Testing gave an average accuracy of 88.85% (247/278) while F-score was 78.85%, based on precision and recall values of 73.4% and 85.3%, respectively.

|  | Actual | |
|---|---|---|
| Predicted | Asphyxia | Normal |
| Asphyxia | 58 | 21 |
| Normal | 10 | 189 |

Table 2: Confusion matrix for Polynomial Kernel experiment showing average accuracy of 88.85% and F-Score of 78.85%.

### B. Radial Basis Function (RBF) kernel experiment

Training converged after 492 iterations with a training set error of 0.0054. Testing gave an average accuracy of 80.93% (225/278) while F-score was 58.26%, based on precision and recall values of 62.7% and 54.4%, respectively.

|  | Actual | |
|---|---|---|
| Predicted | Asphyxia | Normal |
| Asphyxia | 37 | 22 |
| Normal | 31 | 188 |

Table 3: Confusion matrix for Radial Basis Function Kernel experiment showing average accuracy of 80.93% and F-Score of 58.26%.

## V. DISCUSSION

Unlike in most learning problems, the Polynomial Kernel outperforms the Radial Basis Function kernel in this experiment. We suspect it is as a result of the fact that the polynomial kernel considers a combination of features in learning from each training sample; thus, making it consistent with the behavior of the audio wave form where each sample in time is a function of previous samples. This, however, is still being investigated.

## VI. CONCLUSIONS AND FUTURE WORK

Our best result, an average accuracy of 88.85%, slightly surpasses that of the referenced work of Reyes-Galaviz and Reyes-Garcia; thereby supporting the argument that Support Vector Machines actually perform better than Neural Networks on speech-related problems. Our SVM algorithm is also computationally more cost-effective as it converged in a much shorter time (404 epochs) than the referenced experiment. The F-Score, however, goes deeper to show a significantly lower precision in classifying the samples of interest (asphyxia). We believe that this is due to the skewed nature of the training data and thus plan to explore two solution pathways: to obtain more asphyxia data samples for a more balanced training and to apply convenient penalty parameters to the two classes when modeling the learning algorithm.

We believe that the results are promising and indicative of the potentials of a viable solution through more research and access to a larger, diverse dataset. Thus, as part of efforts going forward, we plan to pursue data collection at a local level; with the ultimate objective of moving this research into practical use in the near future.

### ACKNOWLEDGMENT

The *Baby Chillanto Data Base* is a property of *National Institute of Astrophysics and Optical Electronics, CONACYT, Mexico*. We thank Dr. Carlos A. Reyes-Garcia, Dr. Emilio Arch-Tirado and his INR-Mexico group, and Dr. Edgar M. Garcia-Tamayo for their dedication to the collection of the Infant Cry data base.

We also thank Dr. F. K. Opara of the Department of Electrical/Electronic Engineering, Federal University of Technology, Owerri, Nigeria for his supervision and provision of some of the resources that made this research possible.

### REFERENCES

[1] World Health Organization. "The World Health Report," 2005; Chp 5.
[2] United Nations. "We Can End Poverty: Millenium Development Goals and Beyond 2015," Fact Sheet, 2013.
[3] Opportunities for Africa's Newborns: Practical data, policy and programmatic support for newborn care in Africa. Joy Lawn and Kate Kerber, eds. PMNCH, Cape Town, 2006.
[4] O. F. Reyes-Galaviz, and C. A. Reyes-Garcia, "A System for the Processing of Infant Cry to Recognize Pathologies in Recently Born Babies with Neural Networks," SPECOM 2004: 9th Conference, Speech and Computer St. Petersburg, Russia, September 20-22, 2004.
[5] S. Chakrabartty, G. Singh, and G. Cauwenberghs, "Hybrid Support Vector Machine / Hidden Markov Model approach for continuous speech recognition," 43rd IEEE Midwest Symp. On Circuits and Systems, Lansing MI, Aug 8-11, 2000.
[6] J. Mercer, "Functions of positive and negative type and their connection with the theory of integral equations", Philosophical Transactions of the Royal Society A 209 (441–458): 415–446, doi:10.1098/rsta.1909.0016.
[7] H. Chih-Wei, C. Chih-Chung, and L. Chih-Jen, "A Practical Guide to Support Vector Classification," Department of Computer Science, National Taiwan University, Taipei 106, Taiwan, 2010.
[8] G. Yoav, and E. Michael, "SplitSVM: Fast, Space-Efficient, non-Heuristic, Polynomial Kernel Computation for NLP Applications," Proc. ACL-08: HLT, 2008.
[9] Z. Azlee, Y. K. Lee, M. Wahidah, M. Y. Ihsan, and S. Rohilah. "Classification of Infant Cries with Asphyxia Using Multilayer Perceptron Neural Network," Second International Conference on Computer Engineering and Applications, 2010.
[10] M. Brookes, Speech Processing Toolbox for MATLAB, Department of Electrical & Electronic Engineering, Imperial College, Exhibition Road, London SW7 2BT, UK. Home page: http://www.ee.ic.ac.uk/hp/staff/dmb/voicebox/voicebox.html
[11] C. Chih-Chung, and L. Chih-Jen, "LIBSVM: a library for support vector machines," ACM Transactions on Intelligent Systems and Technology, 2:27:1--27:27, 2011. Software available at http://www.csie.ntu.edu.tw/~cjlin/libsvm.
[12] O. F. Reyes-Galaviz, S. D. Cano-Ortiz, and C. A. Reyes-García, "Evolutionary-Neural System to Classify Infant Cry Units for Pathologies Identification in Recently Born Babies," Proceedings of the Special Session MICAI 2008, Pg. 330-335, Eds. Alexander Gelbukh & Eduardo Morales, IEEE Computer Society. ISBN: 978-0-7695-3441-1.